\documentclass[12pt]{iopart}

\usepackage{simplewick}
\usepackage{tikz}

\usepackage{mathrsfs}
\usepackage{amssymb }

\usepackage{epsfig}
\usepackage{verbatim}
\usepackage{multirow}
\usepackage{relsize}

\begin{document}

\title{A new ordering principle in quantum field theory and its consequences}

\author{Jan M. Greben}
\address{CSIR, Pretoria, South Africa (now retired)}
\ead{jmgreben@gmail.com}

\begin{abstract}
The ad-hoc imposition of normal ordering on the Lagrangian, energy-momentum
tensor and currents is a standard tool in quantum field theory (QFT) to eliminate infinite vacuum expectation values (v.e.v.) However, for fermionic expressions these infinite terms are due to anti-particles only. This exposes an asymmetry in standard QFT, which can be traced back to a bias towards particles in the Dirac bra-ket notation. To counter this bias a new ordering principle
(called the $\mathbb{R}$-product) is required which restores the symmetry (or rather duality) between particles and anti-particles and eliminates the infinite v.e.v. While this $\mathbb{R}$-product was already used in a bound-state application, this paper aims to give it a more general foundation and analyze its overall impact in QFT.

For boson fields the particle bias is hidden and the fields must first be expanded into bilinear particle-anti-particle fermion operators. This new representation also leads to vanishing v.e.v.'s and avoids some common technical problems in the quantization of vector fields, while it admits new constant contributions that mimic the Higgs mechanism without unacceptable contributions to the cosmological constant.

Since the $\mathbb{R}$-product does not apply between operators belonging to different space-
time coordinates (e.g. propagators and vacuum condensates), most standard QFT
calculations remain unaffected by this new principle, preserving those quantitative successes. The boson propagator also retains a
standard bosonic form despite the fermionic representation. However, the foundations of QFT are affected strongly as the new
principle suggests that the Standard Model is an effective theory built (partly?) on massless bare quarks.
\end{abstract}


\today


\maketitle

\section{Introduction}
The imposition of normal ordering on the Lagrangian, energy-momentum tensor and currents has become a standard feature in QFT, designed to eliminate the infinities in the vacuum expectation values (v.e.v.) of these dynamical entities.
Although axiomatic components in fundamental theories are generally acceptable, ad hoc features like this should eventually be replaced by a more principled approach or be shown to be a consequence of such principles. As Dirac has noted in the past ``We want to understand how Nature works; to understand the basic ideas which Nature is using, and for that purpose it is not sufficient to have rules giving results in agreement with observation''
(p. 759, \cite{Dirac}). His comments were directed at renormalization, but they are equally valid here.

Let us see how the ad hoc usage of the normal product was defended in the past.
The basic texts on QFT usually start with a discussion of the scalar case which then sets the tone for the rest of QFT and its principles. This order of presentation is an unfortunate one in the current context, as the scalar case does not easily reveal the cause of these unphysical infinities. In the classic texts by Gasiorowicz ([2], p.16) and Bogoliubov and Shirkov ([3], p.103-104, [4], p.76) it is stated that expressions containing products of operators at the same point lead to infinities and must be normal ordered, e.g. in the case of Hamiltonians and currents. Since there is no other obvious cure to the infinity problem in the scalar case, the application of the ad hoc normal ordering recipe has been widely accepted. One has to go to the fermion case to acquire more insight in the source of these infinities.

Schweber ([5], p.228) does discuss the fermion case and notices that the infinite contributions come from anti-particles. However, he associates these with
the charge of the sea of occupied negative energy states. Sakurai ([6], p.151) has a
similar comment, contributing the negative infinite vacuum energy of the vacuum to
anti-particles in the Dirac sea. Neither author appears concerned about the implied
asymmetry between particles and anti-particles. This is unfortunate as this asymmetry holds the key towards a solution of the infinity problem.

In more recent texts by Itzykson and Zuber (\cite{1980Itzykson}, p.111), Kaku (\cite{1993Kaku}, p.67), Peskin and Schroder (\cite{1995Peshkin}, p.22), and Ryder (\cite{1996Ryder}, p.131) the normal product is also introduced for the scalar field,
but the dropping of the infinite vacuum term is justified by arguing that such an energy shift is not measurable. However under general relativity (GR) the absolute energy plays an essential role, so this argument is not a principled one. In addition, the shift argument would not apply to currents as the absolute magnitude of a current is physically important. In discussing the zero point momentum some authors argue that the cancellation between positive and negative momenta in the vacuum integral render the v.e.v. zero, so that no ad hoc prescription is required (e.g. \cite{1980Itzykson}, p.115). This survey shows that the arguments to eliminate the infinite contributions are very diverse (Dirac sea, irrelevant absolute energy scale, ad hoc operational rule to simplify calculations) or opportunistic (we do not need a recipe if there happens to be a cancelation between infinite terms). The discussion is also ambiguous about the nature of zero-point energy: is it physical and is it removed for practical reasons, or is it unphysical and does one have to revise the formulation to ensure that it vanishes? Most theorists seem to think that zero point energy is real and contributes towards the cosmological constant, despite the extreme conflict of this assumption with cosmological observations (a conflict which is not really addressed by introducing cutoffs and renormalization). Even in string theory the role of the ad hoc application of the normal product is still prominent (\cite{1987Green1}, p. 77). In supersymmetry the infinite contributions are considered to be real and the possible cancelation of bosonic and fermionic contributions to the v.e.v. becomes an important motivation for the approach. Clearly, there is no coherent understanding of this problem and a more principled approach is called for.

We will show that by starting with the fermion case, one gets a physical understanding of the reason behind these infinite terms and thereby is able to identify the key towards their removal. This fermionic analysis can then also form the basis for a better understanding and treatment of this phenomenon in the boson case by fermionization of the bosons fields.
\section{Analysis of infinities in the fermion problem}
Most discussions of the application of the ad hoc normal product start with the scalar field. The operator expression for the energy reads:
\begin{equation}
  \label{eq:zeropoint}
\hbar \omega \frac{1}{2}[a^{\dag}(\vec{k})a(\vec{k})+ a(\vec{k})a^{\dag}(\vec{k})]
=\hbar \omega  a^{\dag}(\vec{k}) a(\vec{k})+\frac{1}{2} \hbar \omega \delta_{p}(0)\ .
\end{equation}
After discretization in $\vec{k}$ the delta function is unity and $\frac{1}{2} \hbar \omega$ can be identified as the zero-point energy, corresponding to the lowest harmonic oscillator energy.
Since the v.e.v. of this term in the Hamiltonian is infinite one usually applies the normal product (commuting the creation operators to the left) to eliminate it:
 \begin{equation}
  \label{eq:zeropointx}
\hbar \omega \frac{1}{2}:[a^{\dag}(\vec{k})a(\vec{k})+ a(\vec{k})a^{\dag}(\vec{k})]:
=\hbar \omega  a^{\dag}(\vec{k}) a(\vec{k}).
\end{equation}
Although this procedure is completely ad hoc and is not justified by any physical principle, there is no other obvious way how to eliminate this infinite term. In fact, many theorists nowadays consider this elimination of the infinite term a question of convenience, as they believe that this zero point energy is a real effect, and that the infinity of the v.e.v. can be controlled somehow by renormalization, although the discrepancy with the observed cosmological constant remains totally unexplained.

We now consider the fermion case, taking the discussion by Sakurai (\cite{1967Sakurai}, Eq. 3.364) as a guide. He analyzes the Dirac Hamiltonian:
\begin{equation}
\label{eq:DiracLagrangian}
 {\cal{L}}=\bar{\psi}(x)\left(i \gamma_{\mu}\partial^{\mu} -m\right)\psi(x)
\end{equation}
and after expanding the Hamiltonian in creation and annihilation operators he obtains (\cite{1967Sakurai}, Eq. 3.393):
\begin{eqnarray}
  \label{eq:Hamiltonian}
H=\sum_{s}\int d\vec{p} E_{\vec{p}}
\left( b^{\dag}_{s,\vec{p}}b_{s,\vec{p}}-d_{s,\vec{p}}d^{\dag}_{s,\vec{p}} \right)
\nonumber
\\
=\sum_{s} \int d\vec{p} E_{\vec{p}}
\left( b^{\dag}_{s,\vec{p}}b_{s,\vec{p}}+d^{\dag}_{s,\vec{p}}d_{s,\vec{p}}-\delta_{p}(0) \right)\ .
\end{eqnarray}
Again the v.e.v. of the last term is infinite. In this case the problematic term is exclusively due to anti-particles,
suggesting that the cure to this problem must feature a similar opposing asymmetry. Just like in the scalar case, the standard response to the infinity is to normal order the whole Hamiltonian. However, in this case there is an alternative featuring the asymmetry we anticipated, namely to just normal order the anti-particle term:
\begin{equation}
\label{eq:NormalOrder}
-:d_{s,\vec{p}}d^{\dag}_{s,\vec{p}}:=d^{\dag}_{s,\vec{p}}d_{s,\vec{p}}.
\end{equation}
At the moment this procedure looks just as ad hoc as the general one. However, we can generalize this procedure to longer chains of operators and provide a physical justification for it.
Under this so-called $\mathbb{R}$-product, which was first introduced in \cite {1GrebenQuark}, the order of all anti-particle operators belonging to a \emph{single} space-time variable is reversed between the bra-state on the left, and the ket-state on the right. This operation restores the symmetry (or duality) between particles and anti-particles and as a by-product eliminates the aforementioned infinite terms.

To understand the need for this re-ordering product we have a closer look at the nature of the QFT formulation for Dirac particles. After quantization the Dirac field $\psi$ contains the particle annihilation operator $b_{s,\vec{p}}$ and the anti-particle creation operator $d^{\dag}_{s,\vec{p}}$, while the reverse statements apply to the conjugate field $\bar{\psi}$. The creation of a particle and the destruction of an anti-particle both increase the particle number by one, which is why they belong in one expression. The same is true for the creation of an anti-particle and the destruction of a particle, both of which decrease the particle number by one so that they also belong together. Now this is all pretty obvious, however, its consequences are not. When we now multiply these combinations in order to describe a Lagrangian or Hamiltonian that conserves lepton or baryon number, then we can choose between two options (we only display the operators to make our point):
\begin{equation}
  \label{eq:normal}
\left(\cdots b^{\dag}_{t,\vec{q}}+\cdots d_{t,\vec{q}}\right)
\times\left(\cdots b_{s,\vec{p}}+\cdots d^{\dag}_{s,\vec{p}}\right)
\end{equation}
or
\begin{equation}
  \label{eq:abnormal}
\left(\cdots b_{s,\vec{p}}+\cdots d^{\dag}_{s,\vec{p}}\right)
\times\left(\cdots b^{\dag}_{t,\vec{q}}+\cdots d_{t,\vec{q}}\right).
\end{equation}
Clearly, in QFT one uses the first form. The historical reason is that this corresponds to the form one would use in a non-relativistic theory with only particles: if we have a one particle state (ket vector) $|b^{\dag}_{s,\vec{p}}|0 \rangle$ then the operator $b_{s,\vec{p}}$ cancels that state, while the left-hand operator $b^{\dag}_{t,\vec{q}}$ cancels the final state (bra vector) $\langle 0| b_{t,\vec{q}}|$, thereby selecting the matrix element $A_{ts}(\vec{q},\vec{p})$ in the expansion. However, if the current universe would be dominated by anti-particles then one would need the second form to reach the same goal for anti-particles. This shows that the standard formulation of QFT displays a fundamental asymmetry, due to the historical bias towards particles in non-relativistic formulations. Unfortunately this bias cannot be redressed under the current notational framework of QFT, and there does not seem to exist another mathematical notation which can both capture the usual constraints of QFT, as well as display the required symmetry and duality between particles and anti-particles. Hence, our only alternative appears to be to reverse the order of the \emph{anti-particle} operators between the bra and ket vector, so that the anti-particle operator chains become an exact mirror image of the chains of particle operators. This reversal is effected by the $\mathbb{R}$-product. Had the current universe been dominated by anti-particles then one could have expected a bias towards anti-particles so that the form (\ref{eq:abnormal}) would have been used and the $\mathbb{R}$-product would have to reverse the order of the \emph{particle} operators.

We can also illustrate the need for this product using a less formal, more intuitive pictorial analysis by studying some higher order self-interaction diagrams. Consider an operator expressions like   $b_{\alpha}^{\dag}d_{\beta}^{\dag}d_{\gamma}b_{\delta}$ which is accompanied by a product of amplitudes like $A^{pa}_{\alpha\beta} (x) A^{ap}_{\gamma\delta}(x)$, where the superscript p(a) indicates whether the state is a particle or anti-particle. Other indices of the amplitudes are suppressed as they play no role for the argument.
Let us try to analyze the physical meaning of such a term for a particle state like $|b_{\varepsilon}^{\dag}|0\rangle$. Clearly we cannot use standard Feynman diagrams to represent this term as all associated amplitudes depend on the same $x=(\vec{x},t)$, so all lines would share a single vertex, which makes it difficult to express the physical content of these terms. The Dirac notation implies a sequence of operations between the bra and ket state, which are best represented by a sequence of individual vertices, i.e. the physical mechanisms can best be represented by a diagram that may look like a time ordered Feynman diagram, despite the fact that all space-time points are identical. The particle component of the operator expression, namely $b_{\alpha}^{\dag} b_{\delta}$, has a clear meaning (as one may expect from a formulation that is biased towards particles): a quark in the initial state characterized by a ket vector $|b_{\varepsilon}^{\dag}|0\rangle$, is cancelled by the operator $b_{\delta}$ for $\delta=\varepsilon$, while at the end a quark is recreated to form the final state  $\langle0|b_{\alpha}$ by cancelling the operator $b_{\alpha}^{\dag}$. However, to account for the anti-particle operators we need two vertices: one where particle $\delta$ joins the anti-particle $\gamma$, represented by the amplitude $A^{ap}_{\gamma\delta}(x)$, and another vertex where the anti-particle $\beta$ joins the final particle $\alpha$, represented by the amplitude $A^{pa}_{\alpha\beta} (x)$. Diagrammatically the particle can be represented by a line moving upwards, while the anti-particle moves downwards.
\begin{figure}[h!]
\begin{center}
\includegraphics[width=10cm]{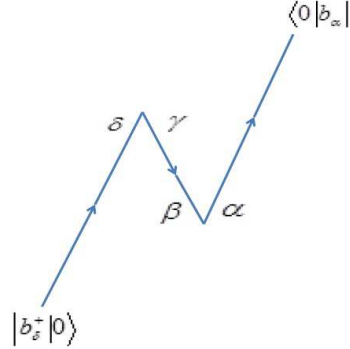}
\end{center}
\caption{Diagram for vertices with same x}
\label{fig:graph}
\end{figure}
These vertices are shown in Fig.\ref{fig:graph}. The $\beta \alpha$ vertex appears first, and the $\delta \gamma$ vertex last. So the operator $d_{\beta}^{\dag}$ in the expression $b_{\alpha}^{\dag}d_{\beta}^{\dag}d_{\gamma}b_{\delta}$ could represent an anti-quark being created from the vacuum followed by its annihilation with $d_{\gamma}$. In terms of our bra-ket representation this would imply that $d_{\beta}^{\dag}$ operates on the ket state, while $d_{\gamma}$ operates on the bra state.
Hence, one needs to reverse the order of the anti-particle operators in order to turn this expression in to a physically acceptable one.
We thus find that the natural representation for this process in the bra ket notation is the reversed operator $-b_{\alpha}^{\dag} d_{\gamma}d_{\beta}^{\dag}b_{\delta}$ (we added a minus sign to respect the anti-commuting nature of fermion operators). The original order with $b_{\alpha}^{\dag}d_{\beta}^{\dag}d_{\gamma}b_{\delta}$ would yield zero on the particle state $|b_{\varepsilon}^{\dag}|0\rangle$, so that this physical interaction term would be eliminated.

On the other hand, the operator $d_{\gamma}b_{\delta} b_{\alpha}^{\dag}d_{\beta}^{\dag}$, which also appears in the expansion, would generate too many terms if left untreated, namely for $\alpha=\delta$ and $\beta=\gamma$. These contributions are not dependent on the character of the initial or final state and thus are unphysical. By similarly reversing the order of the anti-particle operators in this expression, we get the operator $-d_{\beta}^{\dag}b_{\delta} b_{\alpha}^{\dag}d_{\gamma}$, which eliminates the unphysical terms. The physical meaning of this re-ordered operator becomes obvious if we operate on an initial anti-particle state $|d_{\gamma}^{\dag}|0\rangle$. In this case the initial anti-particle state $\gamma$ is followed by the creation of an internal particle line $\alpha$ and the final anti-particle line $\beta$ at the first vertex. The $\alpha$ particle line changes to $\delta$ and meets the initial anti-particle $\gamma$ at the second vertex. Hence, one can only assign a physical meaning to these single variable diagrams if the order of the anti-particle operators is reversed.

Of course, this illustration deals with a very simple diagram. Much more elaborate diagrams can occur, as was the case when we tried to solve the QCD operator field equations for quarks self-consistently \cite{1GrebenQuark}. There long - even infinite - series of operators for a single space-time variable occurred and one had to decide how to deal with such long sequences. However, an exact operator solution of the field equations could be constructed once the $\mathbb{R}$-product was implemented. This product ensured that the important physical interactions were retained, while the prodigious unphysical interactions, which only lead to infinities, were eliminated, just as they were in the example above. The exact operator solution allowed the reduction of the field equations to a finite set of differential equations, which in turn led to a localized solution which could be interpreted as a dressed quark. This illustrates the power of - and need for - this new principle in the case of expectation values or self-consistent QFT calculations of this bound-state character.

Having explained the basic origin of - and need for - the $\mathbb{R}$-product, we now spell out a number of properties of this product. This will also make it plausible why most QFT calculations could be so successful despite the incompleteness of the standard theory. We will discuss later how these properties and results can be applied to the boson case.
\begin{enumerate}
\item \label{1} The general rule for applying the $\mathbb{R}$-product is that for a product of particle and anti-particle operators all belonging to the \emph{same} space-time point we have:
    \begin{equation}
  \label{eq:definition}
\langle f|\mathbb{R}[b_1 \cdots b_n d_1 \cdots d_m]|i\rangle =(-1)^{m(m-1)/2}\langle f| b_1  \cdots b_n d_m \cdots  d_1|i\rangle ,
\end{equation}
where the final expression can be used as a standard operator product without any reference to the space-time variable they belong to. Whether the operators in the chain are creation or annihilation operators is immaterial for the application of this rule (this is the reason for not displaying the creation or annihilation character of these operators in this equation).
\item \label{2}To emphasize the importance of the link to the space-time variable we give an example of a mixed expression where we give the linked space-time point as a (silent) label:
\begin{eqnarray}
  \label{eq:mixed}
\mathbb{R}[d^{(x)}_1 d^{(y)}_2d^{(x)}_3 d^{(y)}_4]=
\mathbb{R}[ d^{(x)}_1 \left\{ d^{(y)}_2,d^{(x)}_3\right\} d^{(y)}_4]-
\mathbb{R}[d^{(x)}_1 d^{(x)}_3 d^{(y)}_2 d^{(y)}_4]=
\nonumber
\\
=\left\{ d_2,d_3 \right\} d_1 d_4-
\mathbb{R}[d^{(x)}_1 d^{(x)}_3 ]\mathbb{R}[d^{(y)}_2 d^{(y)}_4]=
\left\{ d_2,d_3 \right\} d_1 d_4 -d_3 d_1 d_4 d_2
\end{eqnarray}
where we used the fact that the anti-commutator $\left\{ d^{(y)}_2,d^{(x)}_3\right\}$ is not subjected to the
$\mathbb{R}$-product and can be considered a c-number, so that
it can be taken out of the $\mathbb{R}$-product. The $\mathbb{R}$-product prescription can be dropped as soon as none of the operators inside it have the same space-time coordinate or are all particle operators. One can also factorize the $\mathbb{R}$-product into separate $\mathbb{R}$-products for each space-time variable provided the operators are already ordered according to space-time variable.
Once the $\mathbb{R}$-product has been carried out the space-time labels become redundant and can be dropped. If we apply the same reduction without the $\mathbb{R}$-product we would have obtained $\left\{ d_2,d_3 \right\} d_1 d_4 -d_1 d_3 d_2 d_4 $.
\item \label{3} For a common physical expression like $\mathbb{R}[d^{(x)}_1 d^{(x)\dag}_3 d^{(y)}_2 d^{(y)\dag}_4]$  the
result is $d^{\dag}_3 d_1 d^{\dag}_4 d_2 $, i.e. the same as if one had started with the normal ordered expression $:d^{(x)}_1 d^{(x)\dag}_3:$ $:d^{(y)}_2 d^{(y)\dag}_4:= d^{\dag}_3 d_1 d^{\dag}_4 d_2 $.
The equivalence of these prescriptions for such common physical expressions explains why the normal ordered Lagrangian yields correct results in most cases. Notice, that if we apply the normal product we need to specify the nature of the operator (creation or annihilation). This accentuates the very different philosophies of the two prescriptions.
\item \label{4} Under point (\ref{2}) we illustrated the use of the anti-commutator to reduce complex expressions of a mixed form. The anti-commutator referred to operators belonging to different space-time coordinates and could be treated as a c-number. In the case  of a single space-time variable one can also apply such a reduction, however, the result is rather surprising:
\begin{eqnarray}
  \label{eq:commutator}
\mathbb{R}[d_1 \cdots \{d_p,d_{p+1}\}\cdots d_m]
\equiv \mathbb{R}[d_1 \cdots d_p d_{p+1}\cdots d_m+d_1 \cdots d_{p+1} d_p \cdots d_m]
\nonumber
\\
=(-1)^{m(m-1)/2} \left[d_m \cdots d_{p+1} d_p  \cdots  d_1+  d_m \cdots d_p d_{p+1}\cdots  d_1\right]
\nonumber
\\
=(-1)^{m(m-1)/2} d_m \cdots \{d_p, d_{p+1}\} \cdots  d_1
\nonumber
\\
=\{d_p, d_{p+1}\}(-1)^{m(m-1)/2}(-1)^{(m-2)(m-3)/2}\mathbb{R}[d_1  \cdots d_{p-1}d_{p+2}\cdots d_m]
\nonumber
\\
=-\{d_p, d_{p+1}\}\mathbb{R}[d_1  \cdots d_{p-1}d_{p+2}\cdots d_m]
\end{eqnarray}
So the anti-commutator between anti-particle operators linked to the same space-time variable inside an $\mathbb{R}$-product can still be treated as a c-number, but its value features an extra minus sign. In the self-consistent QFT calculations of dressed quarks \cite {1GrebenQuark} we first discovered the need for such minus signs for intermediate anti-particle states. Without these minus signs the equations of motion took on an ugly and unmanageable form, with them they displayed a high degree of symmetry and elegance, and allowed an analytic solution, despite the highly non-linear and strongly coupled nature of the equations. For a long time we suspected a sign error somewhere, however, eventually we found that the needed minus signs could be explained by the identity above. This anti-commutator property is an important tool in the solution of the operator field equations, since it can be applied directly to these equations in $\mathbb{R}$-product form. This facilitates the reduction of these operator equations to manageable differential equations. In perturbative QFT calculations one usually encounters only mixed anti-commutators like $\left\{ d^{(y)}_2,d^{(x)}_3\right\}$ which do not give rise to the extra minus sign, so that the standard techniques remain valid.
\item \label{5} The $\mathbb{R}$-product needs to be applied between the bar and ket vector. So we no longer define the Hamiltonian or Lagrangian itself as normal (or $\mathbb{R}$) ordered, since there may be a chain of operators (all referring to the same space-time point) \emph{between} the ket and bra state vector, and the $\mathbb{R}$-product must be applied to the whole matrix element. So the Lagrangian, Hamiltonian or currents should be defined without any further (normal ordering) prescription.
\item \label{6} Because of the special role of the bra and ket vector, one cannot simply insert bra and ket vectors inside an existing QFT expression. In particular a closure expansion like:
\begin{equation}
  \label{eq:false}
\left \langle f|\mathbb{R}[d_1d_2\cdots d_m]|i\right \rangle =
\sum_j\left \langle f|\mathbb{R}[d_1d_2\cdots d_p|j\right \rangle \left \langle j|\mathbb{R}[d_{p+1} \cdots d_m]|i\right \rangle
\end{equation}
(where the states $\left |j\right \rangle $ form a complete set) is no longer valid in general. Naturally, there are enough circumstances where the insertion works, for example if there are only particle operators or if all anti-particle operators belong to different space-time coordinates (in which case the $\mathbb{R}$-product can be omitted and the resulting chain of operators can be handled in the standard way). But we no longer can rely blindly on this expansion as a basis for general proofs. If one wants to use such an insertion on an anti-particle expression then one should first carry out the $\mathbb{R}$-product, and only then apply closure. Afterwards one can revert to the $\mathbb{R}$-product expression. In this case this would lead to:
\begin{eqnarray}
  \label{eq:true}
\left \langle f|\mathbb{R}[d_1d_2\cdots d_m]|i\right \rangle
\nonumber
\\
=(-1)^{p(m-1)}
\sum_j\left \langle f|\mathbb{R}[d_{p+1} \cdots d_m]|j\right \rangle \left \langle j|\mathbb{R}[d_1 \cdots d_p]|i\right \rangle.
\end{eqnarray}
In practice this expansion will hardly be useful, as closure will usually be applied in cases where the operators belong to different space-time variables.
\item \label{7} The fact that many vacuum matrix elements now yield zero does not prevent the creation of particles from the vacuum (vacuum fluctuations):
\begin{equation}
  \label{eq:quantum fluctuation}
 \langle f|\mathbb{R}[b^{(x)\dag}_1 d^{(x)\dag}_2]|0\rangle \neq 0
\end{equation}
where $ <f |$ is a particle-anti-particle state.
\item \label{8}  In higher order Feynman diagrams long chains of operators might appear. However, since the operators involved usually refer to different space-time variables, the $\mathbb{R}$-product usually does not play a role or has the same effect as the ad hoc normal product. This is one reason why the incompleteness of standard field theory has gone unnoticed in perturbative calculations.
\end{enumerate}

\section{A fermionic expansion of the boson field}
The ad hoc use of normal ordering is usually introduced for scalar fields, as was already noted in the introduction. Since the key to the resolution of the infinity problem in the fermionic case lay in the application of the $\mathbb{R}$-product to anti-particle operators, we can expect that a similar solution would work for bosons. However, in order to apply the $\mathbb{R}$-product to boson fields one must first expand these fields in terms of fermionic operators. While in nuclear structure calculations fermionic expansions of bosons are common (see for example an old paper of ours \cite{Grebenpauli}), in QFT this seems like a drastic step considering the fact that the boson fields are considered very basic and fundamental.
However, it is mainly the operator algebra that is modified and after these operations have been carried out the reduced boson and fermion field entities again act to a large extent as elementary boson and fermion fields. So the fermionization of bosons is not as dramatic a step as one might have anticipated. This will be confirmed later in this section, where we demonstrate that the boson field propagator retains a conventional form under the fermionic expansion.

Boson {-} and in particular scalar {-} fields are generally considered simpler than fermion fields and with the commutators being analogous to the classical Poisson brackets, these fields are often used to introduce field quantization, before the supposedly \emph{more involved} case of Dirac fields is entertained. The current discussion turns this picture around as it suggests that the truly fundamental fields are the fermionic fields which have no clear analogy in classical physics. The boson fields can only be correctly understood after they are expressed in terms of fermionic operators, and so the usual analogies with classical physics are misleading at best. As we will demonstrate below, this fermionic representation of bosons ensures that the usual infinite terms in the v.e.v. do not occur. The vanishing of these vacuum terms implies that the huge discrepancy between theory and observation of the cosmological constant is no longer present. Naturally, one still has to show in more detail that the fermionization of boson fields leads to a consistent theory of bosons and is able to reproduce the successful quantitative results obtained in the standard formulation. The upcoming discussion will also address this point.

In the self-consistent bound-state QFT calculations \cite {1GrebenQuark}, the fermionic expansion of the boson fields was dictated uniquely by the fermionic source term in the quantized boson field equations. These equations automatically ensured that the boson fields had the correct structure and symmetries. The free boson field satisfies a source-less field equation, so one must use other constraints and considerations to determine the nature of a fermionic expansion. After an in depth analysis we found that to lowest order the following expansion satisfies the necessary demands:
\begin{eqnarray}
  \label{eq:boson field}
A^\mu_a (x)=\sqrt{\frac{1}{8SV}}\sum_{\alpha,\beta} \int\frac{d\vec{p}}{\sqrt{2p_0}}
\left[\left(\bar{u}_{\alpha,\frac{\vec{p}}{2}}\gamma^{\mu} \mathbb{O}_a v_{\beta,\frac{\vec{p}}{2}} \right) e^{ipx} b_{\alpha,\frac{\vec{p}}{2}}^{\dag}d_{\beta,\frac{\vec{p}}{2}}^{\dag}\right.
\nonumber
\\
+\left.\left(\bar{v}_{\alpha,\frac{\vec{p}}{2}} \gamma^{\mu} \mathbb{O}_a u_{\beta,\frac{\vec{p}}{2}}\right) e^{-ipx} d_{\alpha,\frac{\vec{p}}{2}}b_{\beta,\frac{\vec{p}}{2}}\right]
\nonumber
\\
+C\sum_{\alpha,\beta} \int d\vec{p}
\left[\left(\bar{u}_{\alpha,\frac{\vec{p}}{2}} \gamma^{\mu} \mathbb{O}_a u_{\beta,\frac{\vec{p}}{2}}\right) b_{\alpha,\frac{\vec{p}}{2}}^{\dag}b_{\beta,\frac{\vec{p}}{2}}\right.
+\left.\left(\bar{v}_{\alpha,\frac{\vec{p}}{2}} \gamma^{\mu} \mathbb{O}_a v_{\beta,\frac{\vec{p}}{2}}\right) d_{\alpha,\frac{\vec{p}}{2}}d_{\beta,\frac{\vec{p}}{2}}^{\dag}\right]
\end{eqnarray}
In this equation $u(v)$ are the free particle (antiparticle) spinors with the discrete quantum number $\beta$, which collectively represents quantum numbers like spin, isospin and colour. The factor $8=2^3$ is a geometric integration constant due to the composition of the boson momentum of two identical contributions from the particle and anti-particle. The inverse volume factor is needed for the proper normalization of the bilinear expansion. The factor $S$ counts the number of discrete states.
The second term is constant and as such represents a new type of contribution to the boson field made possible by the fermionic representation. We will discuss a possible role for this term when we discuss the Higgs field. In the self-consistent quark \emph{bound-state} calculation \cite {1GrebenQuark} terms with this operator structure play an essential role and are not constant.

The matrix elements of the free spinors can be evaluated and we can write more explicitly:
\begin{eqnarray}
  \label{eq:boson explicit}
A^\mu_a (x)=\sqrt{\frac{1}{8SV}}\sum_{\alpha,\beta} \int\frac{d\vec{p}}{\sqrt{2p_0}}
\left[e^{ipx} b_{\alpha,\frac{\vec{p}}{2}}^{\dag}d_{\beta,\frac{\vec{p}}{2}}^{\dag}
+ e^{-ipx} d_{\alpha,\frac{\vec{p}}{2}}b_{\beta,\frac{\vec{p}}{2}}\right]
\nonumber
\\
\times ({\mathbb{O}_a})_{\alpha\beta}\left[ \frac{p^\mu}{2m p_0}\vec{\sigma} \bullet \vec{p}+
\left(
\begin{array}{c}
0\\
\vec{\sigma}-\hat{p}(\vec{\sigma}\bullet \hat{p})\\
\end{array}
\right)^\mu
+\frac{2m}{p_0}
\left(
\begin{array}{c}
0\\
\hat{p}(\vec{\sigma}\bullet \hat{p})\\
\end{array}
\right)^\mu
\right]_{\alpha \beta}
\nonumber
\\
+\frac{C}{m}\sum_{\alpha,\beta} \int d\vec{p}p^{\mu}({\mathbb{O}_a})_{\alpha\beta}{\mathbb{\delta}}_{\alpha\beta}^{spin}
\left[b_{\alpha,\frac{\vec{p}}{2}}^{\dag}b_{\beta,\frac{\vec{p}}{2}}
+d_{\alpha,\frac{\vec{p}}{2}}d_{\beta,\frac{\vec{p}}{2}}^{\dag}\right]
\end{eqnarray}
The fermion field in which we expand the boson field should have all the quantum numbers $\beta$ needed to calculate the matrix element $({\mathbb{O}_a})_{\alpha\beta}$ for all known interactions, i.e. they should be quark spinors. As we want to describe massless photons and gluons these quarks should be bare and massless, so that the parameter $m$ is infinitesimal and should be taken in the limit $m\downarrow 0$ once matrix elements are calculated. Slightly different expressions hold for the scalar Higgs field, which will be discussed at the end of this section.

If we calculate the boson vacuum energy we obtain operator products of the following from (the coefficients are irrelevant for our current considerations):
\begin{eqnarray}
  \label{eq:fermion energy}
  \langle 0|
  \mathbb{R}\left\{\left[\exp(ipx) b_{\alpha,\frac{\vec{p}}{2}}^{\dag}d_{\beta,\frac{\vec{p}}{2}}^{\dag}
+ \exp(-ipx) d_{\alpha,\frac{\vec{p}}{2}}b_{\beta,\frac{\vec{p}}{2}}\right] \right.
\nonumber
\\
\left.\left[\exp(iqx) b_{\gamma,\frac{\vec{q}}{2}}^{\dag}d_{\delta,\frac{\vec{q}}{2}}^{\dag}
+ \exp(-iqx) d_{\gamma,\frac{\vec{q}}{2}}b_{\delta,\frac{\vec{q}}{2}}\right]\right\}
|0\rangle
\nonumber
\\
=\langle 0|\left[e^{ i(p-q)x}\mathbb{R}\{b_{\alpha,\frac{\vec{p}}{2}}^{\dag}d_{\beta,\frac{\vec{p}}{2}}^{\dag}
d_{\gamma,\frac{\vec{q}}{2}}b_{\delta,\frac{\vec{q}}{2}}\}
+e^{ -i(p-q)x}\mathbb{R}\{d_{\alpha,\frac{\vec{p}}{2}}b_{\beta,\frac{\vec{p}}{2}}
b_{\delta,\frac{\vec{q}}{2}}^{\dag}d_{\gamma,\frac{\vec{q}}{2}}^{\dag}\}\right]|0\rangle
\nonumber
\\
=-e^{ i(p-q)x}\langle 0|b_{\alpha,\frac{\vec{p}}{2}}^{\dag}
d_{\gamma,\frac{\vec{q}}{2}}d_{\beta,\frac{\vec{p}}{2}}^{\dag} b_{\delta,\frac{\vec{q}}{2}}|0\rangle
-e^{ -i(p-q)x}\langle 0|d_{\gamma,\frac{\vec{q}}{2}}^{\dag}b_{\beta,\frac{\vec{p}}{2}}
b_{\delta,\frac{\vec{q}}{2}}^{\dag}d_{\alpha,\frac{\vec{p}}{2}}|0\rangle
\end{eqnarray}
The final result displays a very pleasing duality and symmetry. The first term has an anti-particle creation operator on the right, but it yields zero because of the presence of the particle annihilation operator. The second term has a particle creation operator on the right, but it yields zero because of the presence of the anti-particle annihilation operator. Hence, both contributions are zero for {-} what one could call {-} complementary reasons. This must be contrasted with the result for the normal boson representation, when the combination $a^\dag a+a a^\dag$ appears symmetric, but is not: the first term yields nothing, while the last term yields a finite (and after integration infinite) contribution to the v.e.v. The consequence of the fermionic representation is that the bosonic quantum contribution to the cosmological constant also vanishes, thereby resolving the biggest discrepancy in modern physics (see also the discussion in \cite {GrebenCC}). The cosmological constant is thus not determined by QFT processes but must be seen as a constant of Nature, whose presence is a direct consequence of the imposition of the symmetries of general relativity. Its value must be determined by (cosmological) observations. This interpretation of the nature of the cosmological constant also underlies our conformal cosmological theory \cite {GRcosmology}. The elimination of the vacuum energy, better known as the zero point energy, seems to conflict with phenomena like the Casimir effect. However, Jaffe has shown a couple of years ago that such effects can also be explained through regular QFT calculations and do not necessarily require the presence of the zero-point vacuum energy \cite{Jaffe}.

The fact that the v.e.v of an expression like $A_{\mu}(x)A_{\nu}(x)$ (we suppress indices and additional factors which are irrelevant for our argument) vanishes in our approach, may lead to the false impression that our theory is in conflict with established results that feature non-zero v.e.v.'s, such as condensates. The reason for this apparent discrepancy is that many of these so-called v.e.v.'s are derived from limiting procedures involving expressions that originally have distinct space-time variables where the $\mathbb{R}$-product does not apply.
For example, in the derivation of the Gell-Mann-Oakes-Renner relation \cite{Gell-Mann} one takes the limit $x\rightarrow y$ in an expression like $\langle 0|[A_{\mu}(x),A_{\nu}(y)]|0\rangle $, where $A$ is a pseudo-vector amplitude. This procedure leads to matrix elements
\begin{eqnarray}
\delta^{(4)} (x-y)\langle 0|\bar{u}(x)u(y)|0\rangle \equiv\delta^{(4)}(x-y)\langle 0|\bar{u}(x)u(x)|0\rangle ,
\end{eqnarray}
where $u$ is the quark spinor field. It is then stated that the v.e.v. of $\bar{u}u$ is non-zero and can be identified as the pion condensate $\langle 0|\bar{u}u|0 \rangle $. It is clear that under the applied limiting procedure this condensate cannot be identified with the physical vacuum matrix element
$\langle 0|\mathbb{R}\{\bar{u}(x)u(x)\}|0\rangle$, as the latter is zero.
So:
\begin{eqnarray}
\lim_{x\rightarrow y}\langle 0|\bar{u}(x)u(y)|0\rangle \neq\langle 0|\mathbb{R}\{\bar{u}(x)u(x)\}|0\rangle \ .
\end{eqnarray}
This does not mean that the matrix element $\langle 0|\bar{u}(x)u(x)|\rangle 0$ without implied $\mathbb{R}$-product is not useful: it is clear that in these derivations this non-zero matrix element plays a significant role, and therefore is a useful intermediate theoretical expression. Because most standard QFT calculations are based on Feynman diagrams with distinct space-time points, concepts - like condensates - are not affected by our $\mathbb{R}$-product and the vacuum matrix elements based on this limiting procedure retain their usefulness, although denoting them as v.e.v.'s is misleading.

Let us now discuss some of the properties of the new representation for boson fields and indicate where it has advantages over the standard representation. For the electromagnetic field we must replace the matrix element $({\mathbb{O}_a})_{\alpha\beta}$ by $\delta_{\alpha\beta}$  in Eq. (\ref{eq:boson explicit}).
 The field automatically satisfies the condition $\partial_\mu A^\mu_a (x)=0$ as an operator equation (as do all boson fields defined by Eq.(\ref{eq:boson explicit})). Hence we do not need to limit ourselves to a  physical Hilbert space to impose this relationship, as is done in theories like the Gupta-Bleuler model (\cite{6Gupta}, \cite{7Bleuler}).
 The standard way to quantize fields is to decompose them into normal modes where each field component is associated with a different elementary particle. This is clearly the case in our representation as each mode corresponds to (combinations of) different quark states. However, in the standard representation this implies that the four components of $A_{\mu}$ should correspond to four different polarization states. But the photon only has two polarization degrees of freedom, so this creates a problem. Several techniques have been introduced to fix these problems, amongst which so-called gauge fixing which requires the addition of a non-gauge invariant term to the Lagrangian. These problems also affect the Hilbert space which no longer has the character of a quantum state vector space. In our fermionic representation such problems do not appear.
 Also, we do not have awkward commutation rules between the boson operators $a_\mu(\vec{p})$ and $a_\mu^{\dag}(\vec{p})$, which have the wrong sign for $\mu=0$, as all relevant (anti{-}) commutation rules are between fermion field operators for well-defined states.

 If we calculate the electro-magnetic propagator in this representation we find:
\begin{eqnarray}
\label{eq:propagator}
\langle 0|T\left\{A^{\mu} (x)A^{\nu} (y)\right\}|0\rangle =\frac{1}{(2\pi)^3}
\nonumber
\\
\times
\int \frac{d\vec{p}}{2p_0}\{e^{-ip(x-y)}\Theta(x_0-y_0)+e^{ip(x-y)}\Theta(y_0-x_0)\}\left[\frac{p^{\mu}p^{\nu}}{4m^2}-g^{\mu \nu }\right] \ ,
\end{eqnarray}
i.e. there is no trace of the original fermionic representation. The result is a standard massive propagator in the limit $m^2\downarrow 0$. In this derivation we used the identities $\delta(\vec{p}/2-\vec{q}/2)=8\delta(\vec{p}-\vec{q})$ and
$\delta(0)=(2\pi)^{-3}\int d^3x=(2\pi)^{-3}V$.  The spin factor is $S=2$ in the current case.

Finally, we want to discuss the consequences of the higher order terms in the boson field equations for non-Abelean theories such as SU(2) or SU(3). After each iteration of the operator field equations new terms in the solution are introduced which are higher order in terms of bilinear fermion operators. Surprisingly, one can obtain an exact solution of the free non-linear field equations using a single infinite operator $\Lambda$:
\begin{eqnarray}
  \label{eq:boson general}
A^\mu_a (x)=\sqrt{\frac{1}{8SV}}\sum_{\alpha,\beta} \int\frac{d\vec{p}}{\sqrt{2p_0}}
\left[e^{ipx} b_{\alpha,\frac{\vec{p}}{2}}^{\dag}\Lambda d_{\beta,\frac{\vec{p}}{2}}^{\dag}
+ e^{-ipx} d_{\alpha,\frac{\vec{p}}{2}}\Lambda b_{\beta,\frac{\vec{p}}{2}}\right]
\nonumber
\\
\times ({\mathbb{O}_a})_{\alpha\beta}\left[ \frac{p^\mu}{2m p_0}\vec{\sigma} \bullet \vec{p}+
\left(
\begin{array}{c}
0\\
\vec{\sigma}-\hat{p}(\vec{\sigma}\bullet \hat{p})\\
\end{array}
\right)^\mu
+\frac{2m}{p_0}
\left(
\begin{array}{c}
0\\
\hat{p}(\vec{\sigma}\bullet \hat{p})\\
\end{array}
\right)^\mu
\right]_{\alpha \beta}
\nonumber
\\
+\frac{C}{m}\sum_{\alpha,\beta} \int d\vec{p}p^{\mu}({\mathbb{O}_a})_{\alpha\beta}{\mathbb{\delta}}_{\alpha\beta}^{spin}
\left[b_{\alpha,\frac{\vec{p}}{2}}^{\dag}\Lambda b_{\beta,\frac{\vec{p}}{2}}
+d_{\alpha,\frac{\vec{p}}{2}}\Lambda d_{\beta,\frac{\vec{p}}{2}}^{\dag}\right]
\end{eqnarray}
where the operator $\Lambda=\Lambda^{p}\Lambda^{a}=\Lambda^{a}\Lambda^{p}$ is defined as follows:
\begin{eqnarray}
  \label{eq:lambdap}
 \Lambda^{p}=\lim_{n \to \infty }\Lambda^{p}_n; \ \Lambda^{p}_n=\frac{(1-N^{p})}{1}\frac{(2-N^{p})}{2}\cdots \frac{(n-N^{p})}{n}\equiv
 \left(\begin{array}{c}
n-N^{p}\\
n\\
\end{array}\right)
 \\
\Lambda^{a}=\lim_{n \to \infty }\Lambda^{a}_n; \ \Lambda^{a}_n=\frac{(1-N^{a})}{1}\frac{(2-N^{a})}{2}\cdots \frac{(n-N^{a})}{n}
\equiv
 \left(\begin{array}{c}
n-N^{a}\\
n\\
\end{array}\right)
\end{eqnarray}
and
\begin{equation}
  \label{eq:N_operator}
 N^{p}=\sum_{\alpha}\int d^3 p b_{\alpha,\vec{p}}^{\dag} b_{\alpha,\vec{p}}\ ; \qquad
 N^{a}=-\sum_{\alpha}\int d^3 p d_{\alpha,\vec{p}} d_{\alpha,\vec{p}}^{\dag}\ .
 \end{equation}
This is a remarkable result as it constitutes an exact solution to the non-linear field equations. The basic properties used in this derivation are:
\begin{eqnarray}
\label{eq:identities}
b\Lambda^{p}=\Lambda^{p}b^{\dag}=0 \qquad \rm{and} \qquad  d^{\dag}\Lambda^{a}=\Lambda^{a}d=0\ .
\end{eqnarray}
The $\mathbb{R}$-product and the minus sign of the anti-particle anti-commutator Eq.(\ref{eq:commutator}) are absolutely essential for the derivation of this exact solution. This result provides a correction to the operator $\Lambda$ defined in \cite{1GrebenQuark}, which was expressed in terms of products of $(n-N^{p}-N^{a})$, rather than being first factorized in terms of $\Lambda^{p}$ and $\Lambda^{a}$.

The particle and anti-particle $\Lambda$-operators look like projection operators, as:
\begin{equation}
  \label{eq:projection operator}
 \Lambda^{p}\Lambda^{p}=\Lambda^{p};\  \Lambda^{a}\Lambda^{a}=\Lambda^{a}\ ,
 \end{equation}
however, $\Lambda^{p}\Lambda^{a}\neq 0$ so these are not ordinary projection operators. Rather these operators project out only one-particle states. We illustrate this for the case of two-particle ket-states, when $\Lambda$ yields zero in all possible combinations. For a two-particle state we get:
\begin{equation}
  \label{eq:one particle operator}
 b^{\dag}_\delta\Lambda^{p}b_\varepsilon| b^{\dag}_\alpha b^{\dag}_\beta|0\rangle =\delta_{\varepsilon\alpha}b^{\dag}_\delta\Lambda^{p}b^{\dag}_\beta|0\rangle-
 b^{\dag}_\delta\Lambda^{p}b^{\dag}_\alpha b_\varepsilon|  b^{\dag}_\beta|0\rangle=0 \ \quad {\rm{as}} \quad
\Lambda^{p} b^{\dag}=0\ ,
 \end{equation}
while for a two-anti-particle state we get:
\begin{eqnarray}
  \label{eq:one anti particle operator}
\mathbb{R}[ d_\delta\Lambda^{a}d^{\dag}_\varepsilon]| d^{\dag}_\alpha d^{\dag}_\beta|0\rangle
=-d^{\dag}_\varepsilon \mathbb{R}[ \Lambda^{a}]d_\delta d^{\dag}_\alpha d^{\dag}_\beta|0\rangle=
\nonumber
\\
=-\delta_{\delta \alpha} d^{\dag}_\varepsilon \mathbb{R}[ \Lambda^{a}] d^{\dag}_\beta|0\rangle
+d^{\dag}_\varepsilon \mathbb{R}[ \Lambda^{a}] d^{\dag}_\alpha d_\delta d^{\dag}_\beta|0\rangle=
\nonumber
\\
\equiv -\delta_{\delta \alpha} d^{\dag}_\varepsilon \mathbb{R}[ d^{\dag}_\beta \Lambda^{a}] |0\rangle
+d^{\dag}_\varepsilon \mathbb{R}[ d^{\dag}_\alpha\Lambda^{a}]  d_\delta d^{\dag}_\beta|0\rangle=0 \quad {\rm{as}} \quad
d^{\dag} \Lambda^{a}=0 \ .
 \end{eqnarray}
Finally, for  a particle-anti-particle state we get:
\begin{eqnarray}
  \label{eq:boson operation}
\mathbb{R}[ d_\delta\Lambda^{a}\Lambda^{p}d^{\dag}_\varepsilon]| b^{\dag}_\alpha d^{\dag}_\beta|0\rangle=0 \quad
 {\rm{as}} \quad \Lambda^{p}b^{\dag}_\alpha=0 \
 \nonumber
\\
 \mathbb{R}[ b^{\dag}_\delta\Lambda^{a}\Lambda^{p}b_\varepsilon]| b^{\dag}_\alpha d^{\dag}_\beta|0\rangle=0 \quad
 {\rm{as}} \quad \mathbb{R}[ \Lambda^{a}]d^{\dag}_\beta=\mathbb{R}[ d^{\dag}_\beta \Lambda^{a}]=0 \ .
 \end{eqnarray}
The one-particle projection property has important consequences, which were already highlighted in \cite{1GrebenQuark} for the bound-state case. For the current scattering-case it implies that the (energy) expectation values of non-Abelean bosons fields vanish, so that the expectation value differs from the expected value $E=p_0$. This inconsistency might well imply that bosons associated with non-Abelean fields cannot exist as stable physical particles (apart from the fact that they can decay into lighter particles). For QED (an Abelean theory) the operator $\Lambda$ is not required as there are no non-linear terms in the free field equations. Hence, for the photon the demand that the energy expectation value equals $\hbar\omega$ does not lead to this inconsistency. However, even there this consistency condition leads to further demands as the fermionic representation leads to additional unphysical terms in the expectation energy value, which only vanish after further generalizations of the field expansion. This has significant physical consequences which will be discussed in future paper(s).

For the scalar Higgs field the constant term has interesting characteristics and a possible physical role. We write this term as follows
\begin{eqnarray}
\label{eq:Higgs}
\phi^{(c)}(x)=C\sum_{\alpha} \int d\vec{p}
\left[b_{\alpha,\frac{\vec{p}}{2}}^{\dag}\Lambda b_{\alpha,\frac{\vec{p}}{2}}
-d_{\alpha,\frac{\vec{p}}{2}} \Lambda d_{\alpha
,\frac{\vec{p}}{2}}^{\dag}\right].
\end{eqnarray}
If this field is used as an intermediate operator then it simply acts as a constant, as
$\phi^{(c)}(x) b_{\alpha,\vec{p}}^{\dag}\Lambda =8C b_{\alpha,\vec{p}}^{\dag}\Lambda $ and
$\phi^{(c)}(x) d_{\alpha,\vec{p}}\Lambda =8C d_{\alpha,\vec{p}}\Lambda $ . Note that these identities only hold if the operators on the right also refer to the same space-time coordinate $x$. Hence, this constant operator $\phi^{(c)}(x)$ can play the same role as the v.e.v $\langle 0|\phi |0\rangle$ of the Higgs field in the SM Higgs theory by setting $C=\langle 0|\phi |0\rangle/8$ without leading to the usual disturbing infinite vacuum energy contributions. Its other advantage over the standard theory is that one does not have to make an expansion around a classical vacuum expectation value and can maintain the quantum operator character of the field and the field equations. In our opinion classical concepts in QFT should be treated with great care because of the danger of oversimplification. We already saw how the classical analogy with harmonic oscillators was rather misleading.

\section {Summary and historical perspective}
A new ordering principle, embodied by the $\mathbb{R}$-product, can remove the bias towards particles over anti-particles hidden in the standard formulation of QFT. This $\mathbb{R}$-product operates specifically on QFT expressions which contain multiple interactions sharing the same space-time variable. Its application affects vacuum expectation values such as the vacuum energy which no longer feature infinite contributions. The product also enables self-consistent solutions of field equations for self-interacting bound systems, the dressing of single quarks being the prime example \cite{1GrebenQuark}. To apply this new principle to boson fields these fields must be represented in terms of bilinear fermionic operators. Since both fermionic and bosonic QFT energy vacuum expectation values vanish now, the enormous discrepancy between the theoretical estimate of the cosmological constant and observation is no longer present. The fermionic representation of the boson fields also avoids certain technical problems in the quantization of vector fields and creates the possibility of adding constant components to the boson fields. These constant components can take over the role of the v.e.v. of the Higgs field in the SM, thereby maintaining the operator nature of the Higgs field and avoiding the hybrid SM formulation with classical c-number as well as quantum components in the Higgs field.

These results suggest that there is a deeper level below the SM populated by bare massless quarks. Both the boson expansion and the bound-state calculations require the underlying bare quarks to be massless; in the boson case because photons and gluons are massless, in the bound-state case as it would be very inelegant to have unexplained dimensionfull parameters at this fundamental level.
The boson fields then emerge at the SM level, being a result of pointlike particle-anti-particle operator combinations. This picture of a deeper hierarchical level in QFT with fewer elementary particles and parameters is very appealing, but obviously requires further development. Although the average mass of light quarks was predicted very accurately in terms of fundamental constants of Nature \cite{1GrebenQuark},
it is not (yet) clear how this success can be extended to the other generations of quarks. We expect that the Higgs field plays an important role for the higher generations. Our first attempts to incorporate this field in the bound-state calculation has yielded interesting - yet inconclusive - results. The lepton sector is more difficult to model because of the peculiar properties of the weak interactions and the low masses involved. Nonetheless, we hope that the renormalization scheme and the self-consistent bound-state calculations can be stitched together in a consistent scheme, where the bound-state calculations together with the Higgs field furnish the quark and lepton SM masses, while the renormalization procedures can be maintained at the effective SM level.

In this paper we also presented some arguments why the new ordering principle took so long to be discovered. There are some specific reasons for this which are worth mentioning, as they also clarify the role of this principle in relation to other developments. Most importantly, despite its important consequences, there are only a few instances where the need for this principle becomes evident, while in many other cases (such as in the boson case) it is well hidden or can be mimicked by ad hoc procedures like the normal ordering prescription. Even in the case where this principle was first discovered \cite{1GrebenQuark}, its need was established indirectly because it could explain the extra minus signs which were required for diagrams involving anti-particles. Without these minus signs the formulation became unmanageable, while including them led to a highly symmetric, elegant and solvable set of equations. However, once these minus signs were linked to the new ordering principle, the nature and further consequences of this principle became obvious. One may thus ask why a formulation like this was not developed earlier, as any such study would most certainly have revealed the new principle. In order to answer this question we review some aspects of the analysis in \cite{1GrebenQuark}.

In \cite{1GrebenQuark} we developed a self-consistent spatial bound-state formulation within the framework of QCD, where the binding was also described in terms of the underlying QCD theory. This theory was inspired by the MIT quark bag model of nucleons (\cite{Chodos1},\cite{Chodos2}), with the main aim of deriving the confining bag from the non-linear QCD field theory itself, instead of postulating it. The quantum field equations could be solved in terms of the field operators (creation and annihilation operators for the bound wave functions), with two important results for the current context. First, the need for the minus signs mentioned above. Second, after introducing the $\mathbb{R}$-product the field equations yield one exact operator solution which describes a single-particle state, and thus appear to constitute a model of a dressed quark. This suggests that multi-particle bound-state problems cannot really be handled by means of such a spatial QCD formulation with single variable field equations, so that the original intention to derive a common binding potential in a many-body system is not realizable in such a framework. This might explain why most bound-state methods in QFT have continued to use standard S-matrix tools, when in non-relativistic quantum mechanics spatial methods based on the field equations (i.e. the Schrodinger equation) are more common. Examples are the Bethe-Salpeter equations \cite{Bethe} and the Blankenbecler-Sugar approach \cite{Sugar}. Even the currently popular lattice calculations of nucleons \cite{Fodor} use techniques inspired by S-matrix theory. We may thus conclude that via the popular QFT bound-state theories it is difficult to discover the new principle. In the one case where the principle plays an essential role (the dressing of elementary quarks) standard QFT uses renormalization techniques and renormalization parameters to capture the dressing process, so that the study of this case via more dynamical spatial methods without free parameters is also discouraged.

Even in the case where the new principle has a big impact, namely in the cosmological constant problem, an analysis of possible solutions of this problem will not easily guide one towards this new principle. Instead, in supersymmetry one relies on the cancellation between the fermionic and bosonic contributions to the vacuum energy. From our perspective, both contributions are unphysical and vanish individually, so that the imposition of such a cancellation condition does not have any special significance.

In keeping with this paper the SM should be seen as an effective theory, formulated in terms of dressed quarks and bosons that must be represented by pairwise fermion anti-fermion operators. However, for (nearly) all practical purposes the SM fields can be considered as elementary and pointlike, so the SM has proved to be an extremely effective effective theory. Our finding that (light) dressed quarks have a very small radius of about 8 Planck lengths \cite{1GrebenQuark} and that the composite nature of bosons in terms of fermions and anti-fermions does not contradict the pointlike elementary nature of boson fields, goes a long way to explaining this effectiveness. The implied hierarchical structure of particle physics is another example of the hierarchical structures which abound in Nature and make it susceptible to scientific methods.


\section*{References}

\begin{thebibliography}{20}
\bibitem{Dirac} Dirac P M A, 1982 {\it The Requirements of a basic physical theory}, 757-770, Gauge Interactions - Theory and experiment, International School of Subnuclear Physics, Erice, Sicily
\bibitem{1966Gasiorowicz}Gasiorowicz S, 1966 {\it Elementary Particle Physics}, Wiley and Sons, New York
\bibitem{Bogoliubov59} Bogoliubov N N and Shirkov D V , 1959 {\it Introduction to  the Theory of Quantized Fields}, Interscience Publishers, New York
\bibitem{Bogoliubov83} Bogoliubov N N and D V Shirkov, 1983 {\it Quantum Fields}, Benjamin Publishing Comp., Reading, Massachusetts
\bibitem{1961Schweber} Schweber S S, 1961 {\it An Introduction to Relativistic Quantum Field Theory}, Harper and Row, New York
\bibitem{1967Sakurai} Sakurai J J, 1967 {\it Advanced Quantum Mechanics}, Addison-Wesley
\bibitem{1980Itzykson} Itzykson C and Zuber J-B , 1980 {\it Quantum Field Theory}, Mc Graw Hill
\bibitem{1993Kaku} Kaku M, 1993 {\it Quantum Field Theory:  A Modern Introduction}, Oxford University Press
\bibitem{1995Peshkin} Peshkin M E and Schroeder D V, 1995 {\it An Introduction to Quantum Field Theory}, Addison-Wesley
\bibitem{1996Ryder} Ryder L H, 1996 {\it Quantum Field Theory} Cambridge University Press
\bibitem{1987Green1} Green M B, Schwarz J H, and Witten E, 1987 {\it Superstring theory}, Vol. I, Cambridge Monographs on Mathematical Physics
\bibitem{1GrebenQuark} Greben J M, 2012  {\it Quantum Field Theory and the Internal States of Elementary Particles} {arXiv:1208.5406}
\bibitem{Grebenpauli} Greben J M, 1977 {\it Phys. Rev.} {\bf C16}, 776
\bibitem{GrebenCC} Greben J M, 2012 {\it A resolution of the cosmological constant problem}  {arXiv:1209.4734}
\bibitem{GRcosmology} Greben J M, 2010
{\it The Role of Energy Conservation and Vacuum Energy in the Evolution of the Universe}
, Found Sci {\bf 15} 153-176, DOI 10.1007/s10699-010-9172-0, {arXiv:0912.5508}
\bibitem{Jaffe}Jaffe R L, 2005 {\it Phys. Rev.} {\bf D72}, 021301(R)
\bibitem{Gell-Mann} Chanfray G, Ericson M, Kirchbach M. 1994 {\it Mod.Phys.Lett.} {\bf A9}, 279-287
\bibitem{6Gupta} Gupta S, 1950 {\it Proc. Roy. Soc.} A63, 681
\bibitem{7Bleuler} Bleuler K, 1950 {\it Helv. Phys. Acta} 23, 567
\bibitem{Chodos1} Chodos A, Jaffe R L, Johnson K, Thorn C B, and Weisskopf V F, 1974
{\it Phys. Rev.} {\bf D9}, 3471-3495
\bibitem{Chodos2} Chodos A, Jaffe R L, Johnson K and Thorn C B, 1974 {\it Phys. Rev.} {\bf D10},  2599-2604
\bibitem{Bethe} E. E. Salpeter and H. A. Bethe, 1951 {\it Phys. Rev.}{\bf 84}, 1232
\bibitem{Sugar} R. Blankenbecler and R. Sugar, 1966 {\it Phys. Rev.}{\bf 142}, 1051
\bibitem{Fodor} Fodor Z and Hoelbling C, 2012 {\it Light Hadron Masses from Lattice QCD} {arXiv:1203.4789v2 [hep-lat]}

\end{thebibliography}
\end{document}